\documentclass{article}
\usepackage{spconf,amsmath,graphicx}
\usepackage{multirow}
\usepackage{longtable,tabularx}
\usepackage{booktabs}

\usepackage{graphicx}
\usepackage{color}
\usepackage{amsmath}
\usepackage[version=4]{mhchem}
\usepackage{siunitx}
\usepackage{longtable,tabularx}
\newcolumntype{L}[1]{>{\raggedright\arraybackslash}p{#1}}
\newcolumntype{C}[1]{>{\centering\arraybackslash}p{#1}}
\newcolumntype{R}[1]{>{\raggedleft\arraybackslash}p{#1}}
\usepackage{gensymb}
\usepackage[symbol]{footmisc}

\title{Low-cost Measurement of Industrial Shock Signals via Deep Learning Calibration}%
\name{
Houpu Yao$^{1 \star}$~\thanks{$^\star$ These authors contributed equally to this work}, Jingjing Wen$^{2 \star}$, Yi Ren$^{1}$, Bin Wu$^{2}$, Ze Ji$^{3}$
}
\address{$^{1}$ Department of Mechanical and Aerospace Engineering, Arizona State University, United States \\
    $^{2}$ School of Astronautics, Northwestern Polytechnical University, China\\
    $^{3}$ School of Engineering, Cardiff University, United Kingdom}

\begin{document}

\maketitle

\begin{abstract}
Special high-end sensors with expensive hardware are usually needed to measure shock signals with high accuracy. In this paper, we show that cheap low-end sensors calibrated by deep neural networks are also capable to measure high-g shocks accurately. Firstly we perform drop shock tests to collect a dataset of shock signals measured by sensors of different fidelity. Secondly, we propose a novel network to effectively learn both the signal peak and overall shape. The results show that the proposed network is capable to map low-end shock signals to its high-end counterparts with satisfactory accuracy. To the best of our knowledge, this is the first work to apply deep learning techniques to calibrate shock sensors.
\end{abstract}


\section{Introduction}
\label{sec:intro}
Accurate measurement of shock signal is crucial for product design in various industries. Some examples where shock signal matters include, the dropping of electronic devices \cite{wong2008mobiledrop}, the crashing of automobiles \cite{sun2014crashing}, and the landing of aircraft \cite{esmailzadeh1999shimmy}. To verify and validate the design of these products, physical experiments need to be conducted to measure their response under shock loading. However, measuring shock signal with high accuracy can be challenging with traditional accelerometers due to the extreme loading condition especially under the high-g shock environment \cite{younis2006high_g_difficulty}. While existing work in shock signal measurement is mainly based on more reliable but expensive hardware\cite{yuan2017Hopkinson_bar}, in this paper we show that, high-g shock signals can be measured at a much lower cost with low-end sensors after deep learning calibration. 

In shock signal measurement, both the overall signal shape and peak value are of interest to us. The peak value of a shock signal is a very important index in board-level shock test \cite{meng2016mems}, while another important index shock response spectrum (SRS) is decided by the entire shock signal shape \cite{brake2011inverse}. However, the complicated frequency content, short duration, and high magnitude of the shock response pose many difficulties to the accurate measurement of shock signals\cite{lee2010shock}.
Due to the less capable piezoceramic material, the defect in sensor structure design and manufacturing, those low-end shock sensors will become easier to get noise-polluted when measuring shock signals\cite{walter2007history}.
 
Traditional approaches to improve the sensor performance are usually based on designing better but more expensive hardware \cite{zhao2013design_high_g}. Existing research in calibrating shock sensors is mainly focused on making use of Hopkinson bar \cite{yuan2017Hopkinson_bar}. However, instead of directly calibrating the measured signals, Hopkinson bar is primarily used to calibrate the dynamic linearity, sensitivity, and repeatability of the accelerometer. 
Based on Hopkinson bar, researchers have tried to establish the nonlinear relationship of the signal parameters between the sensor output and physical models \cite{wang2017quasi,link2006calibration}. Despite the fact that these methods are very complicated, they are limited to calibrating the peak value and/or pulse width and other important features in the shock signals are ignored. 

Although shock signals are complicated, their internal dynamics are governed by similar physics laws. Inspired by recent progresses in deep learning applied on time series data\cite{song2017attend,wang2017deep}, we believe that deep learning can be a promising tool to find the internal correlations between these shock signals. To the best of our knowledge, the only related work to us is \cite{oh2017force_calib}, which calibrates force sensors with neural networks and numerical simulated data. While both work use data driven approaches to calibrate sensors, we are focusing on high-g shock signals, which brings two extra difficulties: (1) unlike in \cite{oh2017force_calib}, numerical simulated data can be unreliable to serve as references at high-g region due to the extremely nonlinear dynamics \cite{har2003parallel}, (2) we are interested in calibrating both signal peak value and its overall signals shape, which requires better network design to accomplish.

In this paper, we first collect a dataset of shock signals through drop shock test. This dataset includes paired shock signals simultaneously measured by low-end and high-end sensors. Secondly, we propose a network that can accurately map the signals measured by low-end to signals similar to what high-end sensors will produce. Although our network is similar to \cite{song2017attend, oh2017force_calib} for time-series modeling, specific design needs to be made for modeling shock signals. We claim three folds of major contributions in this paper:
\begin{itemize}
  \setlength\itemsep{0pt}
    \item We establish the first dataset for industrial shock signal, which will facilitate the future research in the field of shock signal measurement. 
    \item We propose a novel network which is able to map shock signal to higher fidelity.
    \item We show that data driven approach is promising for measuring complicated shock signals at low cost. 
\end{itemize}

\section{Proposed approach}
In the first part of this section, we describe the acquisition procedure of our shock signal dataset. In the second part, we describe our network structure and how it is trained.

\subsection{Data collection}
\label{subsec:data}
\begin{figure}
    \centering
    \includegraphics[width=0.9\linewidth]{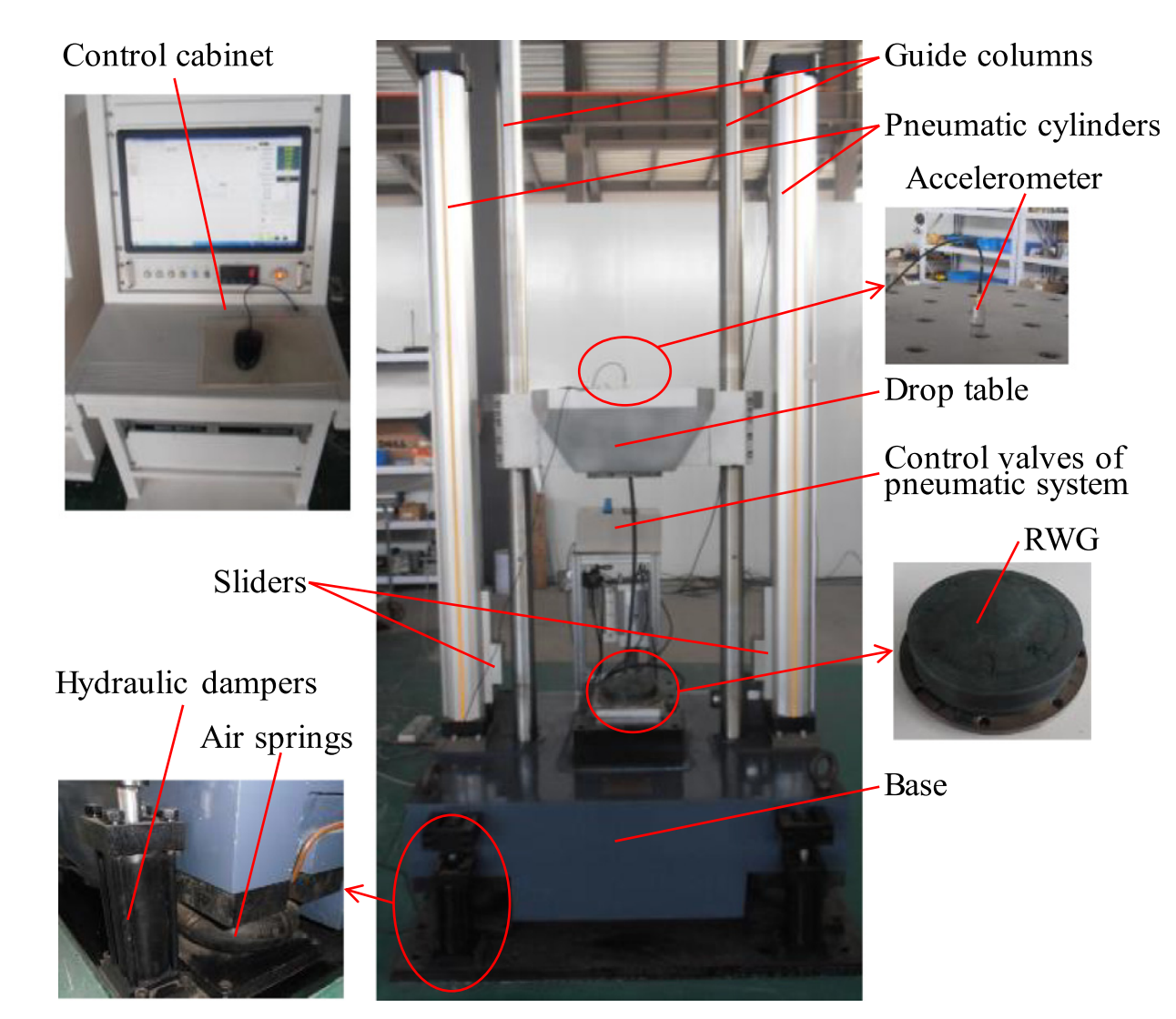}
    \caption{Illustration of the drop test platform. Sensors are marked as ``Accelerometer" mounted on the top of the drop table. Figure adapted from \cite{wen2018nonlinear}. }
    \label{fig:shock_test_platform}
\end{figure}

We conduct drop shock test to collect shock signals. The experiment platform setting is based on our previous work in \cite{wen2018nonlinear}, which is also illustrated in Fig.\ref{fig:shock_test_platform}. The logic behind this drop shock test is simple: the drop table is lifted up and released, it falls freely along the guide columns and collides with the rubber waveform generator (RWG) to produce the shock signal. This shock signal is transmitted to and picked up by the sensors mounted on the top of the drop table. The higher we lift the drop table, the harder the drop table will hit RWG, and the larger the shock signal the system will produce. For more details on the experimental setup, we encourage readers refer to \cite{wen2018nonlinear}.

We used a low-end sensor and a high-end sensor to measure the shock signal simultaneously. Both sensors are attached to the top of the drop table. Low-end sensor is very cheap, but its signal tend to be noisy and can have large error in signal peak value. Meanwhile, high-end sensor is much more expensive but can produce a fairly accurate measurement. We will use the high-end sensor output as the ground-truth. In this study, all sensors are set to have the same sampling frequency of 200kHz. Once a shock signal is generated, these sensors will gather a pair of signals simultaneously with different levels of fidelity. We change the dropping height to obtain different pairs of the measured signal. A total of 660 drop tests are conducted, which leads to 660 collected shock signal pairs. 

\begin{figure}[h]
    \centering
    \includegraphics[width=0.92\linewidth]{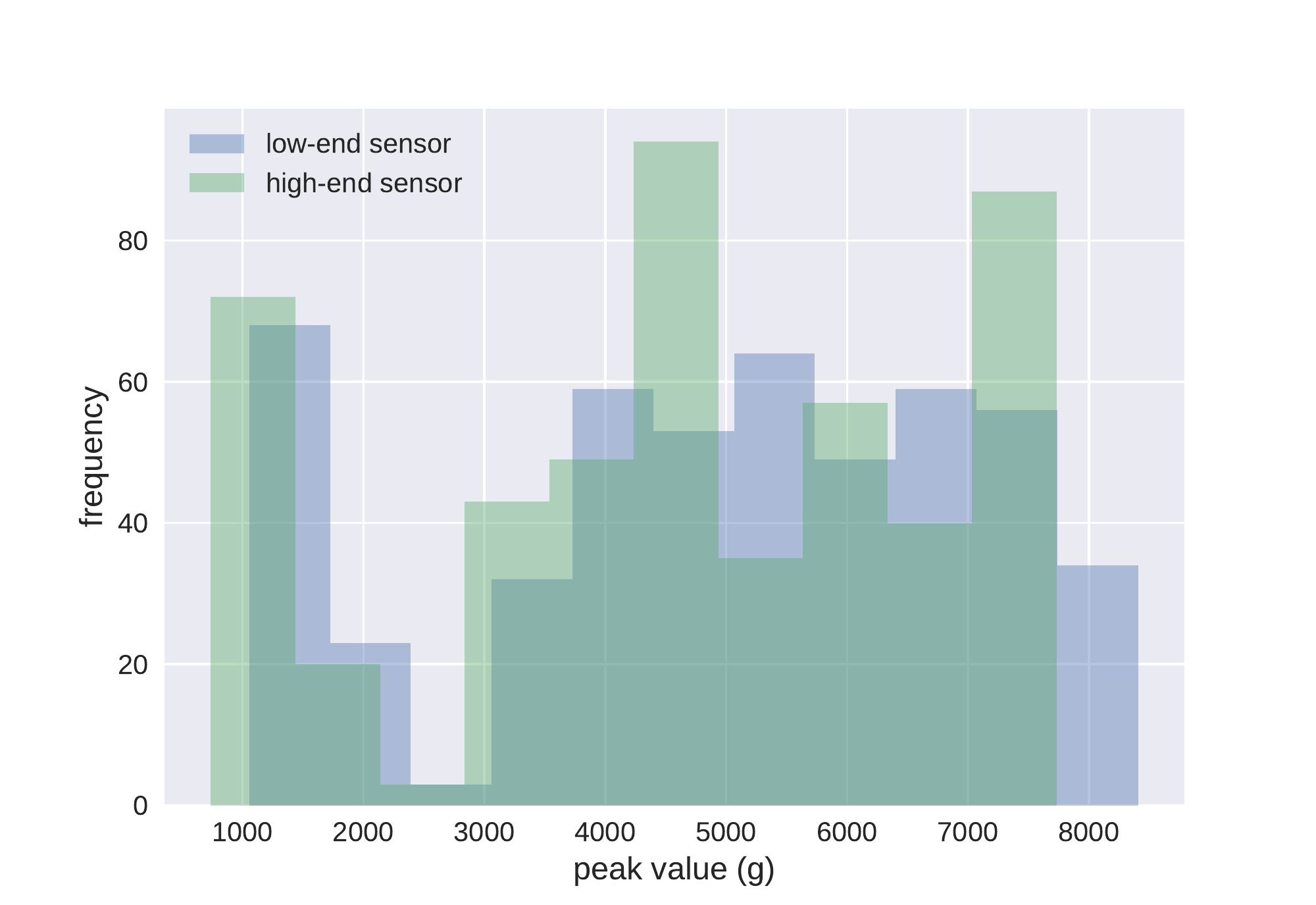}
    \caption{Statistics of the signal peak value in the training set.}
    \label{fig:signal_statistics}
\end{figure}

The raw signal is pre-processed to have equal length. We cut the signals to have an equal duration of 15 ms, with 2.5 ms and 12.5 ms before and after their peak. Considering the sampling rate, each signal has a length of 3000. Samples signals after pre-processing can be seen from Fig.\ref{fig:network}, where the signal to the left and right are produced by low-end sensor and high-end sensor respectively. 
We randomly select and hold back 160 pairs as testing set, while the rest 500 pairs are used to train the network. The distributions of the peak value of the signals from training set are shown in Fig.\ref{fig:signal_statistics}. It can be seen that this dataset covers a wide range of acceleration up to 8,000 g. We will release this dataset to the community for research purpose.

\subsection{Network architecture}
\label{subsec:network}

\begin{figure*}[h]
    \centering
    \includegraphics[width=0.9\linewidth]{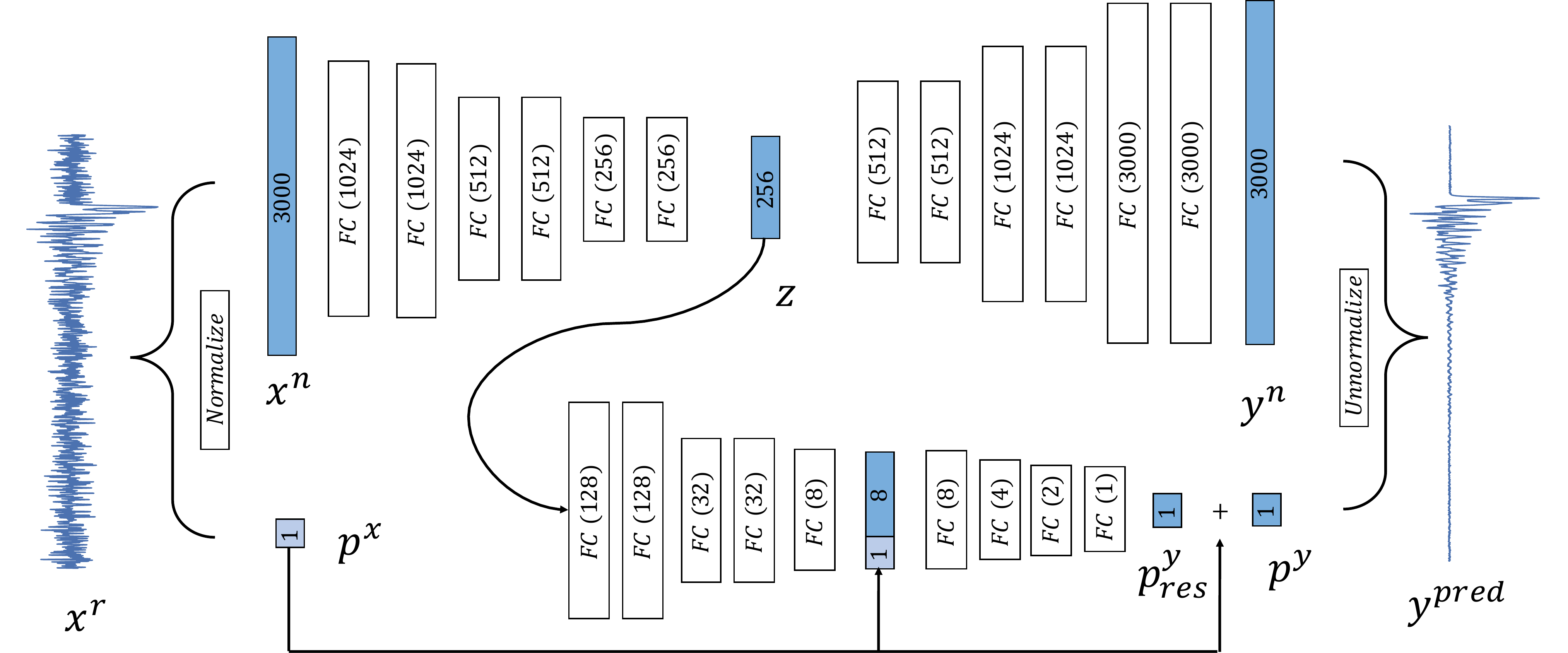}
    \caption{Illustration of proposed network architecture. PPN is the branch at the bottom. Our network takes in a noisy signal $x^r$ as input, and will output a calibrated signal $y^{pred}$. The global feature vector is first further encoded into 8 dimension and then concatenated with the signal peak feature in PPN. Blocks in color denotes tensors, and transparent blocks denote network layers. Numbers in the block corresponds to feature dimensions.}
    \label{fig:network}
\end{figure*}

In this subsection, we propose a novel network architecture to learn the mapping between shock signals obtained with different sensors. The motivation is that, after training, our network can transform the signals produced by low-end sensor to some signals very similar to what high-end sensor will produce. Intuitively, we adopt encoder-decoder style network to learn the mapping between signals produced by different sensors. In addition, as peak value matters for shock signals, we introduce peak prediction network (PPN) to further calibrate the signal peak value.

As illustrated in Fig.\ref{fig:network}, our network has three main parts: encoder, decoder, and PPN. Raw signal $x^r$ is normalized first and its shape $x^n$ and magnitude $p^x$ are feed into the encoder and PPN separately. The normalized signal $x^n$ will be encoded to a 256 dimensional vector $z$, and reconstructed back to a 3000 dimensional vector $y^n$ by the decoder:
\begin{equation}
\begin{split}
        z &= enc(x^n; \theta_1)\\
        y^n &= dec( z; \theta_2)
\end{split}
\end{equation}
where $\theta_1$ and $\theta_2$ are the network parameters for encoder and decoder respectively. 

Because the peak value is crucial for shock signals, we used PPN to further calibrate it. We feed the encoded global information of the normalized signal $z$ as well as the peak value of the input signal $p^x$ to PPN. Its output $p^y_{res}$ is the estimated error between the accurate peak value and the input peak value, which will be added back to $p^x$ to predict and the correct peak value $p^y$:
\begin{equation}
\begin{split}
        p^y_{res} &= ppn(p^x, z; \phi) \\
        p^y &= p^x + p^y_{res}
\end{split}
\end{equation}
where $\phi$ is the network parameter of PPN.


The network loss is composed of two parts:
\begin{equation}
\label{eq:loss}
\begin{split}
    L^{s}(\theta) &= |y^n-y^{ref}|_2 + |y^n-y^{ref}|_{\infty}\\
    L^{p}(\phi) &= |p^{y} - p^{ref}|
\end{split}
\end{equation}
where $L^s$ is the loss to regulate the normalized signal, which is defined on $\theta_1$ and $\theta_2$. $L^p$ is the loss to regulate the peak value of the raw signal, which is defined on $\phi$. The $L_2$ term in $L^s$ loss encourages the transformed signal to have a similar overall shape as the ground truth, and the $L_{\infty}$ term further pushes normalized signal to have correct relative peak value and location. We expect that by minimizing these losses in Eq.\ref{eq:loss}, the network will be able to predict both the overall shape and the peak value of the shock signal well.


\section{Results and Discussion}
\label{sec:result}

\begin{figure*}[!htbp]
    \centering
    \includegraphics[width=0.99\linewidth]{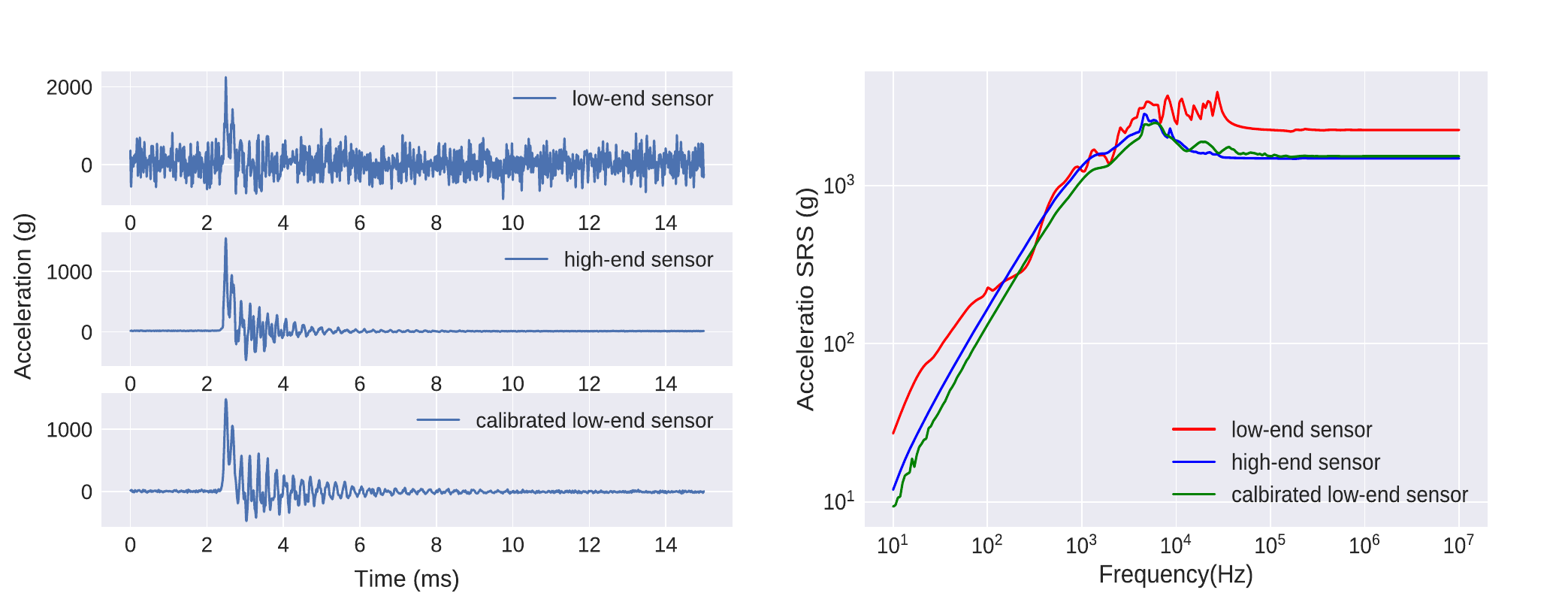}
    \caption{Signal calibration performance of proposed method. Left subplot: visualization in time-domain. Right: visualization in frequency domain with SRS curve. Best viewed in color.}
    \label{fig:rec}
\end{figure*}

As traditional ``hardware" based approaches are costly and the comparison is not quite ``apple to apple", we compare our methods with several possible ``software" based approaches in this section. Candidate ``software" approaches includes low pass filter (LPF), linear regression (LR), and auto-encoder (AE). For LPF, we choose to filter out frequency larger than 5,000 Hz. For AE, we removed the PPN component from the proposed network and kept $L^s$ as training loss.

We propose two metrics $E_{p}$ and $E_{o}$ to measure the calibration performance on peak value and overall signal shape respectively:
\begin{equation}
\begin{split}
    \epsilon_{p} &= \frac{1}{N} \sum _i^N  | \max(y_i^{pred}) -  \max(y_i^{ref}) | /  \max(y_i^{ref})\\
    \epsilon_{s} &= \frac{1}{N} \sum _i^N  \sum_j^M |y_{ij}^{pred} - y_{ij}^{ref}| / \max(y_i^{ref})
\end{split}
\end{equation}
where $M$ and $N$ are signal length (3000 in this case) and the number of signals respectively. $y_i$ is the $i$-th testing signal, and $y_{ij}$ is the signal value of $y_i$ at $j$-th time step. 
As a matter of fact, $\epsilon_{p}$ is the relative prediction error of peak value, and $\epsilon_{s}$ is the relative overall error w.r.t. the peak value. 
Without any treatment, the raw signal from the low-end sensor will have $\epsilon_p$ of 13.5\% and $\epsilon_s$ of 228.6.

The comparison of the calibration performance of these ``software" based approach is listed in Tab.\ref{tab:comp_sft}. It can be seen that the filtering approach will result in a large $\epsilon_p$ as expected, as the signal peak is of high frequency and is hard to be maintained after denoising. In the meantime, all data driven approaches turns out to be effective in calibrating the shock signals. LR is the fastest, but its accuracy on peak prediction is not very satisfactory. While vanilla AE has good performance, proposed network achieves the best result on both $\epsilon_p$ and $\epsilon_s$. As the prediction of the peak value and signal overall shape are decoupled, our proposed network is able to further improve the performance of AE. 

\begin{table}
\centering\renewcommand{\arraystretch}{1.}
\begin{tabular}{|C{0.8cm}|C{1.cm}|C{1.cm}|C{1.cm}|C{1.cm}|C{1.cm}|C{1.cm}|}
\hline
      & Raw  & LPF         & LR       & AE         & Ours        \\ 
\hline
$\epsilon_{p}$  & 13.5\% &   48.8\%    &  7.9\% &  6.9\%    & \textbf{5.7\%}   \\ 
\hline
$\epsilon_{s}$  &228.6    & 138.6    &  44.8  &  37.9   &  \textbf{35.2}      \\ 
\hline
\end{tabular}
\caption{Comparison with other ``software" approaches. }
\label{tab:comp_sft}
\end{table}

An example of the calibrated result from proposed method is shown in Fig.\ref{fig:rec}. The subplot to the left and right shows the calibration effect in time and frequency domain respectively. It can be seen from the left subplot that the signal noise has been largely suppressed after passing through our network. Notably, in the meantime, the signal peak is maintained and even its value has been further calibrated. 
In the right subplot, we adopted the SRS curve to visualize for the calibration effect in frequency domain. SRS is one of the most widely adopted descriptor for shock signals, which can be used to estimate the maximum dynamic response of structures. It is calculated by imposing an excitation in a series of single degree of freedom systems with progressively increasing natural frequency. For more details on its computation we encourage readers refer to \cite{morais2017srs}. It can been seen that while there is a gap between the SRS curve from low-end and high-end signals, the difference is getting very small after our network calibration. This means that the calibrated signal from low-end sensors doesn't have much difference with the signal from high-end sensors in industrial applications.

In order to further understand the effect of each component of our network, we conduct an ablation study on our network structure and loss design.
We first remove the global information $z$ in PPN and find that the $\epsilon_p$ increases to 9.7\%. This means that the global signal shape information actually plays a very important role in predicting the signal peak value. If we only remove $L_{\infty}$ term in $L^s$, it is observed that $\epsilon_p$ will increase by 1\%. This means $L_{\infty}$ term helps better learning of the relative value and location of the shock signal peak. We also tested that removing the ResNet style connection in PPN will increase $\epsilon_p$ by 0.6\%. These result show the effectiveness of our network design.

\section{conclusion}
This is the first time that data driven approaches are introduced to measure shock signals. We designed a novel neural network that is able to calibrate low-end sensors. Results show that, with deep learning calibration, low-end sensors can be used to measure high-g shock signals with satisfactory accuracy. Since industrial signals can be collected in large scale with little effort, we expect that the proposed approach will lower the cost of high-g shock sensor largely in the near future. 


\vfill\pagebreak

\bibliographystyle{IEEEbib}
\bibliography{strings,refs}

\begin{thebibliography}{10}

\bibitem{wong2008mobiledrop}
Ee-Hua Wong, SKW Seah, and VPW Shim,
\newblock ``A review of board level solder joints for mobile applications,''
\newblock {\em Microelectronics Reliability}, vol. 48, no. 11-12, pp.
  1747--1758, 2008.

\bibitem{sun2014crashing}
Guangyong Sun, Fengxiang Xu, Guangyao Li, and Qing Li,
\newblock ``Crashing analysis and multiobjective optimization for thin-walled
  structures with functionally graded thickness,''
\newblock {\em International Journal of Impact Engineering}, vol. 64, pp.
  62--74, 2014.

\bibitem{esmailzadeh1999shimmy}
E~Esmailzadeh and KA~Farzaneh,
\newblock ``Shimmy vibration analysis of aircraft landing gears,''
\newblock {\em Journal of Vibration and Control}, vol. 5, no. 1, pp. 45--56,
  1999.

\bibitem{younis2006high_g_difficulty}
Mohammad~I Younis, Ronald Miles, and Daniel Jordy,
\newblock ``Investigation of the response of microstructures under the combined
  effect of mechanical shock and electrostatic forces,''
\newblock {\em Journal of Micromechanics and Microengineering}, vol. 16, no.
  11, pp. 2463, 2006.

\bibitem{yuan2017Hopkinson_bar}
Kangbo Yuan, Weiguo Guo, Yu~Su, Yunbo Shi, Jingyu Lei, and Hui Guo,
\newblock ``Study on several key problems in shock calibration of high-g
  accelerometers using hopkinson bar,''
\newblock {\em Sensors and Actuators A: Physical}, vol. 258, pp. 1--13, 2017.

\bibitem{meng2016mems}
Jingshi Meng, Stuart~T Douglas, and Abhijit Dasgupta,
\newblock ``Mems packaging reliability in board-level drop tests under severe
  shock and impact loading conditions--part i: Experiment,''
\newblock {\em IEEE Transactions on Components, Packaging and Manufacturing
  Technology}, vol. 6, no. 11, pp. 1595--1603, 2016.

\bibitem{brake2011inverse}
Matthew~Robert Brake,
\newblock ``An inverse shock response spectrum,''
\newblock {\em Mechanical Systems and Signal Processing}, vol. 25, no. 7, pp.
  2654--2672, 2011.

\bibitem{lee2010shock}
Dae-Oen Lee, Jae-Hung Han, Hae-Won Jang, Sung-Hyun Woo, and Kyung-Won Kim,
\newblock ``Shock response prediction of a low altitude earth observation
  satellite during launch vehicle separation,''
\newblock {\em International Journal of Aeronautical and Space Sciences}, vol.
  11, no. 1, pp. 49--57, 2010.

\bibitem{walter2007history}
Patrick~L Walter,
\newblock ``The history of the accelerometer: 1920s-1996-prologue and epilogue,
  2006,''
\newblock {\em Sound \& vibration}, vol. 41, no. 1, pp. 84--90, 2007.

\bibitem{zhao2013design_high_g}
Yulong Zhao, Xiaobo Li, Jing Liang, and Zhuangde Jiang,
\newblock ``Design, fabrication and experiment of a mems piezoresistive high-g
  accelerometer,''
\newblock {\em Journal of Mechanical Science and Technology}, vol. 27, no. 3,
  pp. 831--836, 2013.

\bibitem{wang2017quasi}
Yan Wang, Jinbiao Fan, Jing Zu, and Peng Xu,
\newblock ``Quasi-static calibration method of a high-g accelerometer,''
\newblock {\em Sensors}, vol. 17, no. 2, pp. 409, 2017.

\bibitem{link2006calibration}
A~Link, A~T{\"a}ubner, W~Wabinski, Th~Bruns, and C~Elster,
\newblock ``Calibration of accelerometers: determination of amplitude and phase
  response upon shock excitation,''
\newblock {\em Measurement Science and Technology}, vol. 17, no. 7, pp. 1888,
  2006.

\bibitem{song2017attend}
Huan Song, Deepta Rajan, Jayaraman~J Thiagarajan, and Andreas Spanias,
\newblock ``Attend and diagnose: Clinical time series analysis using attention
  models,''
\newblock {\em arXiv preprint arXiv:1711.03905}, 2017.

\bibitem{wang2017deep}
Yuzhi Wang, Anqi Yang, Xiaoming Chen, Pengjun Wang, Yu~Wang, and Huazhong Yang,
\newblock ``A deep learning approach for blind drift calibration of sensor
  networks,''
\newblock {\em IEEE Sensors Journal}, vol. 17, no. 13, pp. 4158--4171, 2017.

\bibitem{oh2017force_calib}
Hyun~Seok Oh, Gitae Kang, Uikyum Kim, Joon~Kyue Seo, Won~Suk You, and
  Hyouk~Ryeol Choi,
\newblock ``Force/torque sensor calibration method by using deep-learning,''
\newblock in {\em Ubiquitous Robots and Ambient Intelligence (URAI), 2017 14th
  International Conference on}. IEEE, 2017, pp. 777--782.

\bibitem{har2003parallel}
Jason Har and Robert~E Fulton,
\newblock ``A parallel finite element procedure for contact-impact problems,''
\newblock {\em Engineering with Computers}, vol. 19, no. 2-3, pp. 67--84, 2003.

\bibitem{wen2018nonlinear}
Jingjing Wen, Chengwu Liu, Houpu Yao, and Bin Wu,
\newblock ``A nonlinear dynamic model and parameters identification method for
  predicting the shock pulse of rubber waveform generator,''
\newblock {\em International Journal of Impact Engineering}, vol. 120, pp.
  1--15, 2018.

\bibitem{morais2017srs}
OMF Morais and CMA Vasques,
\newblock ``Shock environment design for space equipment testing,''
\newblock {\em Proceedings of the Institution of Mechanical Engineers, Part G:
  Journal of Aerospace Engineering}, vol. 231, no. 6, pp. 1154--1167, 2017.

\end{thebibliography}

\end{document}